\documentclass[a4paper,11pt]{article}
\pdfoutput=1 % if your are submitting a pdflatex (i.e. if you have
\usepackage{jheppub} % for details on the use of the package, please
\usepackage[T1]{fontenc} % if needed

\usepackage{xcolor}
\usepackage{slashed}
\usepackage{amsmath}  
\usepackage{amssymb}  
\usepackage{mathtools}% http://ctan.org/pkg/mathtools
\usepackage{mathrsfs}
\newcommand{\fsl}[1]{\ensuremath{\mathrlap{\mkern-1.7mu\not{\phantom{#1}}}#1}}% \fsl{<symbol>}

\title{Photophilic hadronic axion from heavy magnetic monopoles}

\author{Anton V. Sokolov,}
\author{Andreas Ringwald}

\affiliation{Deutsches Elektronen Synchrotron, Notkestrasse 85, 22607 Hamburg, Germany}

\emailAdd{anton.sokolov@desy.de}
\emailAdd{andreas.ringwald@desy.de}

\abstract{We propose a model for the QCD axion which is realized through a coupling of the Peccei-Quinn scalar field to magnetically charged fermions at high energies. We show that the axion of this model solves the strong CP problem and then integrate out heavy magnetic monopoles using the Schwinger proper time method. We find that the model discussed yields axion couplings to the Standard Model which are drastically different from the ones calculated within the KSVZ/DFSZ-type models, so that large part of the corresponding parameter space can be probed by various projected experiments. Moreover, the axion we introduce is consistent with the astrophysical hints suggested both by anomalous TeV-transparency of the Universe and by excessive cooling of horizontal branch stars in globular clusters. We argue that the leading term for the cosmic axion abundance is not changed compared to the conventional pre-inflationary QCD axion case for axion decay constant $f_a > 10^{12}~\text{GeV}$.}

\begin{document}
	\begin{flushright}
		DESY 21-046\\
	\end{flushright}
	\maketitle
	\flushbottom
	
	\section{Introduction}
	The Standard Model (SM) of particle physics is a very successful theory. Its structure alone predicts many low energy symmetries which were never disproved by any experiment, such as for instance baryon number conservation or time reversal symmetry of quantum electrodynamics. Not all of the possible symmetries of the theory can be however inferred from the structure of the SM, in particular this is the case of time reversal symmetry of Quantum Chromo-Dynamics (QCD). Namely, there is a special free parameter $\bar\theta$ in the SM which indicates whether this symmetry holds. Fortunately, there is also an experimentally accessible observable proportional to $\bar\theta$ -- the neutron electric dipole moment (EDM). While any measured value of this observable would call for some explanation in terms of a more fundamental theory, it is especially challenging that the measurements of the neutron EDM reveal it to be consistent with zero with an unprecedented precision of \mbox{$10^{-26}~e \cdot \text{cm}$}~\cite{Abel:2020gbr}. The question of why QCD is symmetric under time reversal constitutes the core of the so-called strong CP problem. As science aims to explain what we observe, one is tempted to hypothesize a new model where the neutron EDM is constrained to be practically zero. In particular, one of the ideas proposed is to drive this observable to zero dynamically by introducing a new pseudoscalar particle called axion, which is a pseudo Goldstone boson associated to spontaneous breaking of anomalous Peccei-Quinn (PQ) symmetry~\cite{Peccei:1977hh, Peccei:1977ur, Weinberg:1977ma, Wilczek:1977pj}. The great advantage of this mechanism is that the introduction of the axion can naturally solve not only the strong CP problem, but also a much more pressing problem of missing mass in the Universe, i.e. the axion is a perfect candidate for dark matter~\cite{Preskill:1982cy,Abbott:1982af,Dine:1982ah}.
	
	Details of particular axion models can vary. The first axion model proposed, which is the PQWW model~\cite{Peccei:1977hh, Peccei:1977ur, Weinberg:1977ma, Wilczek:1977pj}, identified the axion field with a phase of the Higgs in a two-Higgs-doublet model (2HDM) and was ruled out experimentally soon after the proposal. Then the KSVZ~\cite{Kim:1979if, Shifman:1979if} and DFSZ~\cite{Dine:1981rt, Zhitnitsky:1980tq} axion models were constructed, which were called invisible, because interactions of the corresponding axion particles with the SM are very faint. Such faint they are that even after four decades of exploration the parameter space of these models is still largely terra incognita. Appeal of the invisible models is their simplicity: the DFSZ model exploits the 2HDM just as in the case of the PQWW axion but the axion is now identified with the phase of a new SM-singlet complex scalar field which couples to the Higgses at high energies; while the KSVZ model exploits coupling of a new SM-singlet complex scalar field, the phase of which is identified with the axion, to a new heavy quark. Over the years, there have been attempts of constructing axion models which would be more "visible" than the DFSZ and KSVZ models, however it always turned out that simplicity was to be sacrificed. For example, in the clockwork axion model~\cite{Farina:2016tgd}, in order to get an enhancement of the axion-photon coupling by six orders of magnitude compared to the KSVZ model, one has to introduce at least 13 new scalar fields. A similar enhancement by six orders of magnitude in all couplings to SM particles is achievable in the $\mathbb{Z}_{\mathcal N}$ axion model~\cite{Hook:2018jle,DiLuzio:2021pxd}, but it requires ${\mathcal N}=45$ copies of the SM. Although quite non-minimal from the theory side, such an enhancement would allow one to explain some uneven astrophysical observations concerning cooling of the horizontal branch stars in globular clusters~\cite{Ayala:2014pea} and anomalous TeV-transparency of the Universe~\cite{DeAngelis:2008sk, Horns:2012fx}, not to mention that such photophilic axions can be well probed experimentally in the nearest future. 
	
	Motivation of this work is then to build a photophilic axion model which possesses the advantages listed above, but which involves a minimal number of new fields and representations. We show that this can be done by introducing a SM-singlet complex scalar field which couples to magnetic monopoles at high energy. In particular, we proceed as follows. First, in section~\ref{sec2}, we briefly review the current status of Abelian as well as non-Abelian magnetic monopoles and discuss the charge quantization condition. Then, in section~\ref{sec3}, we describe our axion model and outline the solution to the strong CP problem it provides, proceeding to section~\ref{sec4}, where we compute the low energy effective axion Lagrangian. Finally, we discuss phenomenology and cosmology of the model in section~\ref{sec5}. 
	
	There is yet another motivation for our study, which is to broaden the current understanding of axion models and of possible implementations of the PQ mechanism. The model we present has qualitative features which no other axion model possesses: in particular, it has an increased sensitivity to the structure of infrared (IR) QCD, its axion-gluon coupling is not automatically standard, and it predicts magnetic monopoles. Besides new experimental prospects, it can well provide a basis for novel insights on the structure of the ultraviolet (UV) theory, which is finally to give a more fundamental description of nature than the SM does.
	
	\section{Abelian and non-Abelian magnetic monopoles}\label{sec2}
	
	As it was shown by Dirac~\cite{Dirac:1931kp}, the observed quantization of charge in electrodynamics can be elegantly explained by adding a magnetic monopole to the theory. The consistency condition for a theory with both electric and magnetic currents is:
	\begin{equation}\label{dirac}
	e g = 2 \pi n\, ,\;\, n \in \mathbb{Z} \, ,
	\end{equation}
	where $e$ is the elementary electric charge and $g$ is the magnetic charge of the monopole. Local Lagrangian quantum field theory (QFT) of Dirac magnetic monopoles was constructed later by Zwanziger~\cite{PhysRevD.3.880}. In order to obtain a consistent local theory, both electric and magnetic four-potentials  ($A_{\mu}, B_{\mu}$) had to be introduced which are sourced by magnetically and electrically charged particles, respectively. It turned out that the resulting Lagrangian possesses an $SO(2)$ symmetry, which rotates charges ($e, g$) and four-potentials ($A_{\mu}, B_{\mu}$) in the electric-magnetic plane, which is broken to a $\mathbb{Z}_2$ symmetry exchanging electric and magnetic quantities in the full quantum theory: $(e, g) \rightarrow (g, -e)$ and $(A_{\mu}, B_{\mu}) \rightarrow (B_{\mu}, -A_{\mu})$. Moreover, the total gauge group of this theory is $U_{\text{e}}(1) \times U_{\text{g}}(1)$, electric charges transforming in a representation of the "electric" $U_{\text{e}}(1)$ group, while magnetic charges transform in a representation of the "magnetic" $U_{\text{g}}(1)$ group. Due to the condition~\eqref{dirac} the theory is essentially non-perturbative, the asymptotic Dyson series being not well-defined and the corresponding Feynman diagrams as well as the Lagrangian itself losing Lorentz covariance. Of course, failure of our perturbative techniques does not mean that the theory is by itself inconsistent and indeed, it was formally shown by Brandt, Neri and Zwanziger~\cite{PhysRevLett.40.147, PhysRevD.19.1153} using the  path-integral approach that observables of the Zwanziger theory are Lorentz invariant if the Dirac condition~\eqref{dirac} is satisfied. Note that this analysis was performed both in the case where electric and magnetic particles are complex scalars as well as in the case where they are all Dirac fermions. In the case where electric and magnetic particles include both fermions and scalars, it is known that the Zwanziger effective theory is not enough for an adequate description of the low energy phenomena, the most famous example being proton decay due to the Rubakov-Callan effect~\cite{Rubakov:1981rg, Callan:1982au}. Although spectacular, such violations of the decoupling principle will not concern us in this work, since we will exploit solely fermionic electric and magnetic charges in our model.

	With the advent of the Standard Model (SM) of particle physics, the Dirac condition~(\ref{dirac}) was extended~\cite{Englert:1976ng} to include all possible types of magnetic charges $\vec{Q}_{M_i}$ in the theory:
	\begin{equation}\label{qgen}
	\exp\left( i \sum\limits_{i=1}^{r} \vec{Q}_{M_i} \, \vec{\mathcal{H}} \right) = 1\, ,
	\end{equation}
	where $\mathcal{H}_k \equiv e_k \cdot h_k$ are Cartan generators of the Lie algebra $\mathcal{G}$ of rank $r$ of the gauge group multiplied by the corresponding electric charges $e_k$. In case of a non-Abelian gauge theory, $e_k$ are equal to the gauge couplings of the theory. For the SM, at low energies, we have $\mathcal{G} = \mathfrak{su}(3) \oplus \mathfrak{u}(1)$, which means that a magnetically charged particle has generally Abelian as well as non-Abelian magnetic charges. In this theory, the minimal magnetic charge corresponding to the electromagnetic subgroup, is still $g = 2\pi/e$, although there are now fractionally charged quarks. The reason is that quarks interact strongly with the monopole that has a color magnetic charge, compensating the would-be observable phase which results from the electromagnetic interaction. In particular, for a down-type quark the quantization condition~(\ref{qgen}) can be written as:
	\begin{equation}\label{dquark}
	\xi g_s t_3 + \zeta \sqrt{3}\,g_s t_8 - \frac{e}{3}\, g = 2\pi\cdot \mathrm{diag} \left( n_1, n_2, n_3 \right),
	\end{equation}\label{SMqucond}
	where $\xi, \zeta \in \mathbb{R}$, $n_1, n_2, n_3 \in \mathbb{Z}$, $t_3 = \lambda_3 / 2$, $t_8 = \lambda_8 / 2$; $\lambda_a$~are Gell-Mann matrices; $g_s$ is the strong coupling. Coexistence of a monopole with charged leptons requires $eg = 2\pi m, \, m \in \mathbb{Z}$. Then Eq.~(\ref{dquark}) can be solved with respect to the coefficients $\xi, \zeta$:
	\begin{equation}\label{coeff}
	\xi = \frac{2\pi}{g_s}\cdot \left(2n_1 + n_3+m \right) , \,\,\, 
	\zeta = -\frac{2\pi}{g_s}\cdot \left( n_3 + \frac{m}{3}   \right) 
	%b = -\frac{2\pi}{3 g_s}\, m - \frac{2\pi}{g_s}\, n_3 
	\,. 
	\end{equation}
	Note that the quantization condition for up-type quarks is satisfied automatically as long as Eq.~(\ref{dquark}) holds, for their electric charges differ by one elementary charge $e$ from those of the down-type quarks. One can see that $m=1$, which corresponds to the minimal Dirac magnetic charge, is still possible, although magnetic monopole must carry non-Abelian magnetic charge as well. The latter is not necessary in the case $m=3$ where viable solutions include $\xi = \zeta = 0$, which means vanishing non-Abelian magnetic charge. 
	
	Having discussed QFT of the Abelian magnetic monopoles and the generic quantization condition pertinent to both Abelian and non-Abelian magnetic charges, let us outline the status of the theory of the latter. First, we note that the condition~\eqref{qgen} can be expressed in a simple way using the language of the Lie group theory. In particular, Goddard, Nuyts and Olive~\cite{Goddard:1976qe} showed that the condition~\eqref{qgen} in a theory with gauge group $G$  can be regarded as a one-to-one correspondence between the magnetic charges of monopoles in this theory and the weights of the Langlands dual gauge group $G^{V}$, which is now also known as the GNO group. For example, the gauge group of electromagnetism is self-dual in this sense: $\left( U(1) \right)^{\!V} = U(1)$; and the GNO group corresponding to the gauge theory of QCD can be inferred from the following identity: $ \left( SU(3)/\mathbb{Z}_3 \right) ^{\!V} = SU(3)$. Based on the derived relation between magnetic charges and the dual gauge group $G^{V}$, which is completely analogous to the relation between electric charges and the gauge group $G$, Goddard, Nuyts and Olive suggested that magnetic monopoles of a gauge theory with a group $G$ generally transform in the representations of the group $G^{V}$. The above conjecture, known as the GNO conjecture, obviously holds in the case of the Abelian group $G = U(1)$, for which the Zwanziger theory discussed earlier in this section can be constructed. The GNO conjecture for the non-Abelian monopoles, in its stronger form known as the Montonen-Olive conjecture~\cite{Montonen:1977sn}, has recently been proven by Kapustin and Witten~\cite{Kapustin:2006pk} for a twisted $\cal{N}$ $=4$ supersymmetric Yang-Mills (YM) theory. In this work we assume that the GNO conjecture holds for the gauge theory of QCD as well, inspired by the findings of Hong-Mo, Faridani and Tsun~\cite{Chan:1995xr} that the classical (nonsupersymmetric) YM equations possess a generalized dual symmetry similar to the electric-magnetic $\mathbb{Z}_2$ symmetry of the Zwanziger theory mentioned above. Let us also note, that although non-Abelian magnetic charges are often introduced as emergent from spontaneous breaking of some larger gauge symmetry, the results by Goddard, Nuyts and Olive do not depend on such a construction and can be as well stated for generic magnetic monopoles defined in the fiber bundle framework of Wu and Yang~\cite{Wu:1975es}. We will thus consider magnetic particles as fundamental in this work, leaving aside the questions concerning their possible inner structure. 
	
	For concreteness, in the next sections we limit ourselves to the two minimal magnetic charge assignments: a pure Abelian magnetic monopole with a charge $6\pi / e$ and a non-Abelian color-magnetic monopole with an Abelian magnetic charge $2\pi / e$, which correspond respectively to the cases $m=3$ and $m=1$ discussed after  Eq.~(\ref{coeff}). For the non-Abelian case, we will consider only magnetic charges transforming in the fundamental representation of $SU(3)$ with the coupling constant $2\pi / g_s \, $, bearing in mind that the higher representation GNO monopoles are unstable due to the Brandt-Neri-Coleman analysis~\cite{Brandt:1979kk, Coleman:1982cx}.

	\section{Solution to the strong CP problem}\label{sec3}
	
	Suppose there exist a vector-like fermionic magnetic monopole $\psi = {\psi}_L + {\psi}_R $ which transforms under an anomalous PQ symmetry $U(1)_{PQ}$~\cite{Peccei:1977hh, Peccei:1977ur} and a complex scalar field $\Phi$ which breaks the PQ symmetry spontaneously at some high energy scale $v_a$. As discussed in the previous section, we consider minimal magnetic charge assignments corresponding either to the Abelian (electromagnetic) monopole or to the non-Abelian (color-magnetic) one. In the former case we assume that $\psi$ transforms in a fundamental representation of the QCD gauge group, i.e. it is a new quark. As far as we do not consider the electromagnetic interaction, such model with a new quark is an exact analog of the KSVZ axion model and thus it provides a solution to the strong CP problem in the same way the KSVZ model does. The aim of this section is then to show that the model with the non-Abelian color-magnetic monopole solves the strong CP problem as well. The high-energy Lagrangian in this case includes the following terms:
	\begin{equation}\label{startlag}
	\mathcal{L} \; \supset \; i\bar{\psi} \gamma^{\mu} \partial_{\mu} \psi + \bar{\psi} \gamma^{\mu} C_{\mu} \psi + y \left( \Phi \, \bar{\psi}_L {\psi}_R + \text{h.c.}\, \right) - \lambda_{\Phi} \left( \left| \Phi \right|^{2} -\frac{v_a^2}{2} \right)^{\!\! 2},
	\end{equation}
	where $C_{\mu}$ is a connection on a GNO group $SU(3)$ multiplied by the corresponding magnetic coupling: $C_{\mu} = g_{\text{m}}\, t_a\, C^{ a}_{\mu}$. In the broken phase, there exists a pseudo Goldstone boson $a$ (axion), which can be introduced via the polar decomposition of the PQ scalar field $\Phi = \frac{1}{\sqrt{2}} \left( v_a + \sigma \right) \cdot \exp{(-ia/v_a)}$ near the vacuum. Let us dispose of the axion dependence in the Yukawa term by performing a chiral rotation of the fermions $\psi \rightarrow \exp{(ia\gamma_5 / 2v_a)}\cdot \psi$. Omitting the terms containing a heavy radial field $\sigma$, one then obtains:
	\begin{equation}\label{lag}
	\mathcal{L} \; \supset \; i\bar{\psi} \gamma^{\mu} \partial_{\mu} \psi + \bar{\psi} \gamma^{\mu} C_{\mu} \psi + \frac{y v_a}{\sqrt{2}}\, \bar{\psi} \psi - \frac{\partial^{\mu} a}{2v_a}\, \bar{\psi} \gamma_{\mu} \gamma_5 \psi + \mathcal{L}_{ \text{F}} ,
	\end{equation}
	where $\mathcal{L}_{ \text{F}}$ is a Fujikawa contribution coming from the  transformation of the fermion measure in the path integral, i.e. the density of the index of the Dirac operator $\gamma^{\mu} D_{\mu} = \gamma^{\mu} ( \partial_{\mu} - C_{\mu} )$. By the Atiyah-Singer index theorem, the latter is equal to the characteristic class of the GNO group bundle, so that:
	\begin{equation}\label{Fuji}
	\mathcal{L}_{\text{F}} = - \frac{a}{16\pi^2 v_a} \, \textrm{tr} \, C_{\mu \nu} \tilde{C}^{ \mu \nu},
	\end{equation}
	where $C_{\mu \nu}$ is the curvature of the GNO group connection and $\tilde{C}_{\mu \nu} = \epsilon_{\mu \nu \lambda \rho}\, C^{\lambda \rho}/2\,$, $\, \epsilon_{0123} \equiv 1$.
	
	In order to see that such a model provides a solution to the strong CP problem, we invoke Abelian gauge fixing introduced by 't Hooft~\cite{tHooft:1981bkw}.
	In the Abelian gauges there arise singularities corresponding to effective color magnetic currents which result in the violation of the non-Abelian Bianchi identities (VNABI)~\cite{Bonati:2010tz}. The time reversal violating term of the QCD action can then be expanded as follows:
	\begin{eqnarray}\label{thetaterm}
	\mathcal{S}_{\text{QCD}} \; \supset \; \frac{\bar\theta g_s^2}{32\pi^2} \int d^4 x \, \sum_{a =1}^8 G_{\mu \nu}^a \tilde{G}^{a\, \mu \nu} \; = \; \frac{\bar\theta g_s^2}{32\pi^2} \, \times \qquad \qquad \qquad \qquad \qquad \qquad \quad && \nonumber \\
	\left\lbrace \int d^4x\,\, \epsilon_{\mu \nu \lambda \rho}\, \partial^{\mu} \! \! \! \sum_{a,b,c\, = 1}^8 \!\! \left(  A_{a}^{\nu}\, G_{a}^{\lambda \rho} - \frac{1}{3}\, g_s f_{abc}\, A_a^{\nu} A_b^{\lambda} A_c^{\rho} \right) - 2\int d^4x \, \sum_{a =1}^8 A^{a}_{\nu} \left( D_{\mu} \tilde{G}^{\mu \nu} \right)_{\!a} \right\rbrace  , \qquad \; &&
	\end{eqnarray}
	where $G_{\mu \nu}^a$ ($A_{\mu}^a$) are components of the non-Abelian field strength tensor $G_{\mu \nu}$ (four-potential $A_{\mu}$) of QCD, $f_{abc}$ are $\mathfrak{su}(3)$ structure constants, $\bar\theta$ is QCD vacuum angle, $\tilde{G}_{\mu \nu}^a = \epsilon_{\mu \nu \lambda \rho}\, G^{a\, \lambda \rho}/2\,$. 
	
	Let us consider the first term on the right-hand side of Eq.~(\ref{thetaterm}). Since all the singularities characteristic of the Abelian 't Hooft gauges arise in the diagonal part of the gluon field, i.e. in the components $A_{\mu}^3$ and $A_{\mu}^8$, the terms of the integrand which contain solely off-diagonal fields can be safely integrated with the use of the Stokes theorem:
	\begin{eqnarray}\label{1term}
	\int d^4x\,\, \epsilon_{\mu \nu \lambda \rho}\, \partial^{\mu} \!\! \! \sum_{a,b,c\, = 1}^8 \!\! \left(  A_{a}^{\nu}\, G_a^{\lambda \rho} - \frac{1}{3}\, g_s f_{abc}\, A_a^{\nu} A_b^{\lambda} A_c^{\rho} \right) = \qquad \qquad \qquad \qquad \qquad \quad && \nonumber \\
	\int d^4x\,\, \epsilon_{\mu \nu \lambda \rho}\, \partial^{\mu} \sum_{\alpha =3,8}\, \sum_{b,c\, = 1}^8 \left(  A_{\alpha}^{\nu} \mathscr{G}_{\alpha}^{\lambda \rho} + 2\, g_s f_{\alpha bc}\, A_{\alpha}^{\nu} A_b^{\lambda} A_c^{\rho}  \right) + 
	\int\limits_{\Omega_{\infty}} dS^{\mu}\, \mathfrak{K_{\mu}} \left[ A_{\text{off-diag}} \right] \, , \qquad \quad &&
	\end{eqnarray}
	where $\mathscr{G}_{\alpha}^{\mu \nu} =\, \partial^{\mu}\! A^{\nu}_{\alpha}\, - \, \partial^{\nu}\! A^{\mu}_{\alpha}$ ($\alpha = 3, 8$) are Abelian field strength tensors. 
	As it is derived both from theoretical considerations~\cite{Kondo:2014sta} and lattice calculations~\cite{Amemiya:1998jz}, in the Abelian gauges off-diagonal gluons obtain finite mass, which means that the functional $\mathfrak{K_{\mu}} \left[ A_{\text{off-diag}} \right]$ vanishes at the surface at infinity, $\Omega_{\infty}$. For the same reason the integrand in Eq.~(\ref{1term}) proportional to $ \partial^{\mu} ( A_{\alpha}^{\nu} A_b^{\lambda} A_c^{\rho} ) $ is restricted to arbitrarily small surfaces around the singularities after application of the Stokes theorem and finally integrates to zero due to regularity of the off-diagonal fields. 
	
	Equation~(\ref{thetaterm}) can now be rewritten in the following way:
	\begin{eqnarray}\label{almo}
	\int d^4 x \, \sum_{a =1}^8 G_{\mu \nu}^a \tilde{G}^{a\, \mu \nu} \, = \, \int d^4 x \sum_{\alpha = 3,8} \mathscr{G}_{\mu \nu}^{\alpha} \tilde{\mathscr{G}}^{\alpha\, \mu \nu} + \qquad \qquad \qquad \qquad \qquad && \nonumber \\
	2\int d^4 x \left( \sum_{\alpha =3,8} A_{\alpha \, \nu} \partial_{\mu} \tilde{\mathscr{G}}_{\alpha}^{\mu \nu} - \sum_{a =1}^8 A_{a\, \nu}\left( D_{\mu} \tilde{G}^{\mu \nu} \right)_{\!a} \right) \! . \; &&
	\end{eqnarray}
	Let us show that the VNABI, $D_{\mu} \tilde{G}^{\mu \nu}$, is diagonal in color space, so that the second row in Eq.~(\ref{almo}) equals to zero. First, note that the only contribution to VNABI comes from singularities, where topological defects associated with the monopoles hamper commutation of partial derivatives, so that in the expression for a commutator of covariant derivatives,
	\begin{eqnarray}\label{commut}
	\left[ D_{\rho}, D_{\lambda}\right] \, = \, -iG_{\rho \lambda} \, + \, \left[ \partial_{\rho}, \partial_{\lambda}\right] \,,
	\end{eqnarray}
	the second term on the right does not vanish. After taking advantage of Eq.~(\ref{commut}) and Jacobi identities for partial as well as covariant derivatives, the expression for \mbox{VNABI} can be simplified~\cite{Suzuki:2017lco}:
	\begin{equation}\label{vnabi}
	D_{\mu} \tilde{G}^{\mu \nu} =\,\, \frac{1}{2}\, \epsilon_{\mu \nu \rho \lambda} \left[ D^{\mu}, G^{\rho \lambda} \right]  =\,\, \frac{1}{2}\, \epsilon_{\mu \nu \rho \lambda}  \left[ \partial^{\rho}, \partial^{\lambda} \right] A^{\mu} = \partial_{\rho}  \tilde{\mathscr{G}}^{\rho \nu} ,
	\end{equation}
	where in the last step only diagonal gluons survive. One can see that the diagonal form of VNABI is ensured by its linearity in the $A_{\mu}$ field. We note that the second term on the right-hand side of Eq.~(\ref{thetaterm}) is then nothing but a manifestation of the Witten effect~\cite{Witten:1979ey}: QCD monopoles are dyons with color electric charges proportional to the vacuum angle $\bar\theta$.
	
	Due to the identities Eqs.~(\ref{almo}) and (\ref{vnabi}) the CP violating term of the QCD Lagrangian reduces in the Abelian gauges to
	\begin{equation}\label{thetaAb}
	\frac{\bar\theta g_s^2}{32\pi^2}  \sum_{\alpha = 3,8} \mathscr{G}_{\mu \nu}^{\alpha} \tilde{\mathscr{G}}^{\alpha\, \mu \nu} \, ,
	\end{equation}
	which involves now only Abelian four-potentials. By the analogous transformation of the Fujikawa contribution~\eqref{Fuji} to the axion Lagrangian~\eqref{lag}, i.e. choosing the same Abelian gauge in the GNO gauge group, one obtains the term for the interaction of the axion with the Abelian dual four-potentials:
	\begin{equation}\label{FujiAb}
	\mathcal{L}_{\text{F}} = - \frac{a g_{\text{m}}^2}{32\pi^2 v_a}  \sum_{\alpha = 3,8}  C_{\mu \nu}^{\alpha} \tilde{C}^{\alpha \, \mu \nu},
	\end{equation}
	where the axion field is assumed to be constant and homogeneous, since this is a vacuum expectation value of it which is a key to the PQ mechanism. Now that we have abelianized the relevant terms, we are in the realm of the Zwanziger theory, so that the electric and magnetic four-potentials can be related due to the dual $\mathbb{Z}_2$ symmetry\footnote{The existence of the SM quarks -- given the absence of their magnetic partners -- obviously violates the electric-magnetic symmetry of this $U(1)^2$ Zwanziger-like theory. However, these quarks are known to be massive. This means they have no relevance for the instanton vacuum effects which are responsible for the generation of the $\bar\theta$-term}, $C_{\mu \nu}^{\alpha} = \tilde{\mathscr{G}}^{\alpha}_{\mu \nu}$, which yields: 
	\begin{equation}
	\mathcal{L}_{\text{QCD}} \; \supset \; \frac{v_a g_s^2 \, \bar\theta + ag_{\text{m}}^2}{32\pi^2 v_a}  \sum_{\alpha = 3,8} \mathscr{G}_{\mu \nu}^{\alpha} \tilde{\mathscr{G}}^{\alpha\, \mu \nu} \, .
	\end{equation}
	Physically, this is just an instantiation of the fact that the $U(1)$ electric and magnetic fields enter the expressions~\eqref{thetaAb} and~\eqref{FujiAb} symmetrically, as products $\vec{E} \cdot \vec{B}$. The standard PQ mechanism is now in order: redefinition of the pseudo Goldstone axion field $a \rightarrow a - v_a \bar\theta \, g_s^2 / g_{\text{m}}^2 $ absorbs the $\bar\theta$-term into the axion-gluon term and subsequent application of the  Vafa-Witten theorem~\cite{Vafa:1984xg} ensures $\langle a \rangle = 0$. The strong CP problem is thus solved.
	
	\section{Calculation of the effective Lagrangian}\label{sec4}
	
	Let us return to the original Lagrangian~\eqref{startlag} and derive the corresponding low energy physical phenomena. For that, we use a linear decomposition of the PQ field, $\Phi = \left( v_a + \sigma + i a \right) \! / \sqrt{2}$, where $a$ is a pseudo Goldstone axion field.\footnote{$\sigma$ and $a$ introduced here are different from the fields denoted by the same letters in Sec.~\ref{sec3}, but there should be no confusion, since different notations are restricted to different sections.} Below the PQ scale, the field $\sigma$ decouples and we are left with the Lagrangian involving axion and heavy monopoles:
	\begin{equation}
	\mathcal{L} \; \supset \; i\bar{\psi} \gamma^{\mu} \partial_{\mu} \psi + \bar{\psi} \gamma^{\mu} C_{\mu} \psi + \frac{y v_a}{\sqrt{2}}\, \bar{\psi} \psi + \frac{i y}{\sqrt{2}}\, a\bar{\psi} \gamma_5 \psi \, ,
	\end{equation}
	where $C_{\mu}$ now also includes the electromagnetic four-potential and corresponds in general to the connection on either of the two GNO gauge groups, Abelian or non-Abelian, discussed in the end of Sec.~\ref{sec2}.
	The aim of this section is to integrate out the heavy field $\psi$. The beauty of the pseudoscalar interaction is that in this case the calculations can be done exactly, without the need of perturbative expansion in the coupling constant. In order to get an effective Lagrangian at low energy we use the proper time method~\cite{Schwinger:1951nm} developed by Schwinger. The effective pseudoscalar current is
	\begin{eqnarray}\label{pseucur}
	J_a = i \left\langle \, C \left| \bar{\psi} ( x ) \gamma_5 \psi ( x ) \right| C \right\rangle = -\, i\, \frac{y v_a}{\sqrt{2}} \int_{0}^{\infty} ds\, e^{-is y^2 v_a^2/2}\; \textrm{tr} \left[ \langle x | \gamma_5 e^{-i \hat{H} s} | x \rangle \right],
	\end{eqnarray}
	with the proper time Hamiltonian
	\begin{equation}\label{ham}
	\hat{H} = - \left( \fsl{\hat{p}} - \fsl{C} ( \hat{x} ) \right)^2 = -\left( \hat{p}_{\mu} - C_{\mu}(\hat{x}) \right)^2 + \frac{1}{2}\,\sigma^{\mu \nu} C_{\mu \nu} ( \hat{x} ) \, ,
	\end{equation}
	where $\sigma_{\mu \nu} = \frac{i}{2} \left[ \gamma_{\mu}, \gamma_{\nu} \right]$, $\; C_{\mu \nu} = \partial_{\mu} C_{\nu} - \partial_{\nu} C_{\mu} + \left[ C_{\mu}, C_{\nu} \right] \,$ and $\fsl{a} \equiv a_{\mu} \gamma^{\mu}$.
	
	First, our goal is to evaluate the matrix element entering Eq.~(\ref{pseucur}), which modulo $\gamma_5$ denotes the probability amplitude of returning to the same point $x^{\mu}$ in Minkowski space after proper time~$s$. Note that since we are interested in the phenomenology at energies much less than the PQ scale $v_a$ and the fluctuations of heavy fields $\psi$ are possible only at the spatial and temporal extent $\sim v_a^{-1}$, external gauge fields in the following calculation can be considered constant. Our calculation of the pseudoscalar current then closely follows that performed by Schwinger~\cite{Schwinger:1951nm}, although we are considering generic non-Abelian GNO group connection instead of the electromagnetic four-potential. 
	We solve the Heisenberg equations of motion in a constant field $C_{\mu \nu}$,  
	\begin{align}
	& \frac{d\hat{\pi}_{\mu}}{ds} = i \left[ \hat{H}, \hat{\pi}_{\mu} \right] = 2\, C_{\mu \nu } \hat{\pi}^{\nu}\, , \\[4pt]
	& \frac{d\hat{x}_{\mu}}{ds} = i \left[ \hat{H}, \hat{x}_{\mu} \right] = 2\, \hat{\pi}_{\mu}
	\, ,
	\end{align}
	and find the generalized momentum $\hat{\pi}_{\mu} = \hat{p}_{\mu} - C_{\mu}$ and position $\hat{x}_{\mu}$ as a function of proper time $s$:
	\begin{align}\label{pimu}
	&\hat{\pi}_{\mu} (s) = e^{2s\, C_{\mu \nu}}\, \hat{\pi}^{\nu} (0)\, ,\\
	\label{xmu} & \hat{x}_{\mu} (s) = \hat{x}_{\mu} (0) + 2 \left( C^{\mu \lambda} \right)^{-1}\cdot e^{s C^{\lambda \nu}} \sinh{sC_{\nu \rho}}\cdot   \hat{\pi}^{\rho} (0)\, .
	\end{align}
	Next, with the use of Eqs.~(\ref{pimu}) and (\ref{xmu}) we rewrite the Hamiltonian~(\ref{ham}) in terms of position operators $\hat{x}_{\mu} (s)$ and $\hat{x}_{\mu}(0)$:
	\begin{equation}\label{modham}
	\hat{H}\, \supset \; -\, \frac{1}{4} \left( \sinh{sC_{\kappa \lambda}} \right)^{-1} C_{\lambda \nu} C^{\nu \rho} \left( \sinh{sC^{\rho \sigma}} \right)^{-1} \times \left[ \hat{x}_{\kappa} (s), \hat{x}^{\sigma} (0) \right] \; + \;  \frac{1}{2}\, \sigma_{\mu \nu} C^{\mu \nu} \, , 
	\end{equation}
	leaving only the terms that do not vanish after taking the matrix element $\langle x(0) | \hat{H} |x(s) \rangle \propto \langle x(0) | x(s) \rangle$. Note that the exponents coming from Eqs.~(\ref{pimu}), (\ref{xmu}) contract into the identity matrix due to antisymmetricity of the field strength tensor $C_{\mu \nu}$. The commutator in Eq.~(\ref{modham}) is easily calculated with the help of Eq.~(\ref{xmu}) and canonical commutation relations. Since the Hamiltonian~(\ref{modham}) is a generator of proper time translations, one can write now a differential equation for the sought-after matrix element:
	\begin{equation}
	i\partial_s \langle x(0) | x(s) \rangle = \langle x(0) | \hat{H} | x(s) \rangle = \langle x(0) | x(s) \rangle \times \left\lbrace \frac{i}{2}\, C_{\mu \nu} \coth{sC^{\mu \nu}} \; + \; \frac{1}{2}\, \sigma_{\mu \nu} C^{\mu \nu} \right\rbrace  .
	\end{equation}
	The solution is:
	\begin{equation}
	\langle x(0) | x(s) \rangle = A\, \frac{\textrm{pf}\, C_{\alpha \beta}}{\textrm{pf} \sinh{sC_{\alpha \beta}}} \cdot \exp \left( -\frac{is}{2}\sigma_{\mu \nu} C^{\mu \nu} \right),
	\end{equation}
	where $A = -i/(4\pi)^2$ is an integration constant which is calculated by matching with the elementary case of vanishing field strength $\mathscr{G}^{\alpha} = 0$. A skew-symmetric four-by-four matrix has two pairs of opposite sign eigenvalues, which we denote as $\pm \Lambda_1$, $\pm \Lambda_2$ in the particular case of $C_{\alpha \beta}$. The trace entering Eq.~(\ref{pseucur}) can be now rewritten in the following form:
	\begin{equation}
	\textrm{tr} \left[ \langle x | \gamma_5 e^{-i \hat{H} s} | x \rangle \right] = -\, \frac{i}{16\pi^2}\, \textrm{tr}_c \left[ \frac{\Lambda_1 \Lambda_2}{\sinh{s \Lambda_1} \sinh{s \Lambda_2}} \times \textrm{tr}_{\gamma} \left\lbrace \gamma_5 \exp \left( -\frac{is}{2}\sigma_{\mu \nu} C^{\mu \nu} \right) \right\rbrace \right] ,
	\end{equation}
	where we have explicitly separated traces over colour ($\textrm{tr}_c$) and spinor ($\textrm{tr}_{\gamma} $) indices. Sums over the spinor indices can be performed using simple algebraic relations, namely $(\sigma_{\mu \nu} C^{\mu \nu})^2 = 8I_1 + 8i\gamma_5 I_2$, $\gamma_5^2 = 1$, $\textrm{tr}\, \gamma_5 = \textrm{tr}\, \sigma_{\mu \nu} = \textrm{tr}\, \gamma_5 \sigma_{\mu \nu}  = 0$, where $I_1 \equiv C_{\mu \nu} C^{\mu \nu}/4$, $I_2 \equiv \epsilon_{\mu \nu \lambda \rho}\, C^{\mu \nu} C^{\lambda \rho}/8\, $:
	\begin{equation}
	\textrm{tr}_{\gamma} \left\lbrace \gamma_5 \exp \left( -\frac{is}{2}\sigma_{\mu \nu} C^{\mu \nu} \right) \right\rbrace = 4i\, \mathrm{Im} \cosh{sX} = 4 \sinh{\left( s \, \frac{X+X^*}{2} \right)} \sinh{\left( s\, \frac{X-X^*}{2} \right) }\, ,
	\end{equation}
	where $X \equiv i\sqrt{2}\cdot \sqrt{I_1 + i I_2}$. Quite conveniently, by solving the characteristic equation for the matrix $C_{\alpha \beta}$, which has the form $\Lambda^4 + 2I_1 \Lambda^2 - I_2^2 = 0$, one can infer that
	\begin{equation}
	\Lambda_1 = \frac{X+X^*}{2}\, , \quad \Lambda_2 = \frac{X-X^*}{2}\, ,
	\end{equation}
	and the overall expression for the current simplifies into 
	\begin{equation}\label{finja}
	J_a = \frac{i y v_a}{4\sqrt{2}\pi^2}\, \textrm{tr}_c ( I_2 ) \int_{0}^{\infty} ds\, e^{-is y^2 v_a^2/2} = \frac{1}{16\sqrt{2}\pi^2 y v_a}\, \epsilon_{\mu \nu \lambda \rho}\, \textrm{tr}_c \! \left( C^{\mu \nu} C^{\lambda \rho} \right) .
	\end{equation}
	
	Finally, we calculate the trace over color indices and expand in terms of the electromagnetic and color gauge fields:
	\begin{spreadlines}{1ex}
		\begin{equation}	
		J_a = \frac{1}{8\sqrt{2}\pi^2 y v_a} \times 
		\begin{dcases}
		-\, 3g_1^2 F_{\mu \nu} \tilde{F}^{ \mu \nu} + \frac{g_m^2}{2}\,  G^a_{(d)}{{}_{\mu \nu}} \tilde{G}^{a\, \mu \nu}_{(d)} \, , \\
		-\, 3g_2^2 F_{\mu \nu} \tilde{F}^{ \mu \nu} + \frac{g_s^2}{2}\,  G_{\mu \nu}^a \tilde{G}^{a\, \mu \nu} \, ,
		\end{dcases}
		\end{equation}
	\end{spreadlines}
	where summation over $a=1\dots 8$ is implied; $g_1 = 2\pi / e$ and $g_2 = 6\pi / e$ -- we separated the two cases discussed in the end of Sec.~\ref{sec2}, corresponding to the stable non-Abelian monopole and the minimal Abelian one, respectively. We also introduced notation for the dual gluon fields $G^a_{(d)}{{}_{\mu \nu}}$, which are components of the connection on the color GNO group, and expressed the dual electromagnetic field strength tensor in terms of the conventional one using $F_{(d)}{{}_{\mu \nu}} = \tilde{F}_{ \mu \nu}$, which obviously holds for constant fields, since the vacuum Maxwell equations are dual-invariant. The effective axion Lagrangian is then given by the following expression:
	\begin{spreadlines}{1ex}
		\begin{equation}\label{leffageneral}
		\mathcal{L}_{\text{eff}}\; \supset \; \frac{y}{\sqrt{2}}\, a J_a \; = \; \frac{a}{16\pi^2 v_a} \times
		\begin{dcases}
		-\, \frac{3}{4\alpha^2}\, e^2 F_{\mu \nu} \tilde{F}^{ \mu \nu} + \frac{1}{8\alpha_s^2}\, g_s^2 \, G^a_{(d)}{{}_{\mu \nu}} \tilde{G}^{a\, \mu \nu}_{(d)} \, , \\
		-\, \frac{27}{4\alpha^2}\, e^2 F_{\mu \nu} \tilde{F}^{ \mu \nu} + \frac{1}{2}\, g_s^2\,  G_{\mu \nu}^a \tilde{G}^{a\, \mu \nu} \, , 
		\end{dcases}
		\end{equation}
	\end{spreadlines}
	where we introduced the fine-structure constant $\alpha = e^2/ 4\pi$ and its QCD analogue $\alpha_s = g_s^2/ 4\pi$.

	\section{Phenomenology}\label{sec5}

	Let us introduce the axion decay constant $f_a = 4\alpha_s^2 v_a$ ($ f_a = v_a $), for the case of the non-Abelian (Abelian) monopole. Using the dual symmetry of a Zwanziger-like theory describing diagonal gluons, we obtain the relation between the magnetic and electric $U(1)$ field strength tensors $\mathscr{G}^{\alpha}_{(d)} = \tilde{\mathscr{G}}^{\alpha}$ ($\alpha =3,8$). The effective Lagrangian Eq.~(\ref{leffageneral}) can then be rewritten in the following form:
	\begin{spreadlines}{1ex}
		\begin{equation}\label{efflag}
		\mathcal{L}_{\text{eff}}\; \supset \; \begin{dcases}
		-\, \frac{1}{4} \left(g^0_{a\gamma} \right)_1 a  \, F_{\mu \nu} \tilde{F}^{ \mu \nu}\, -\, \frac{a g_s^2}{32\pi^2 f_a}\; G_{\mu \nu}^a \tilde{G}^{a\, \mu \nu} + \mathcal{L}_{\text{off}}\, , \\
		-\, \frac{1}{4} \left(g^0_{a\gamma} \right)_2 a \,  F_{\mu \nu} \tilde{F}^{ \mu \nu}\, +\, \frac{a g_s^2}{32\pi^2 f_a}\; G_{\mu \nu}^a \tilde{G}^{a\, \mu \nu} \, ,
		\end{dcases}
		\end{equation} 
	\end{spreadlines}
	where
	\begin{spreadlines}{1ex}
		\begin{equation}\label{g0}
		g^0_{a\gamma} = \begin{dcases}
		3\alpha_s^2 / \left( \pi \alpha f_a  \right) \, ,\\
		27 / \left( 4 \pi \alpha f_a \right) \, ,
		\end{dcases}
		\end{equation}
	\end{spreadlines}
	is a coupling of axion to photons. For convenience, we separated some axion-gluon interactions into $\mathcal{L}_{\text{off}}$, which is given by the following expression:
	\begin{eqnarray}\label{off}
	\mathcal{L}_{\text{off}} && =\, \frac{a g_s^2}{32\pi^2 f_a} \times \left( G_{\mu \nu}^{a} \tilde{G}^{a\, \mu \nu} - \sum_{\alpha =3,8} \mathscr{G}_{\mu \nu}^{a} \tilde{\mathscr{G}}^{a\, \mu \nu} \, + \, \left( A \rightarrow A_{(d)} \right) \right) = \qquad \qquad \nonumber \\
	-\, \frac{g^2_s\, \partial^{\mu} a}{16\pi^2 f_a} && \! \! \epsilon_{\mu \nu \rho \lambda} \times
	\left\lbrace \sum_{i,j,k\, \in \, \mathcal{I}_{\text{off}} } \!\! \left(  A_i^{\nu}\, \partial^{\, \rho} A_i^{\lambda} + \frac{1}{3}\, g_s\,  f_{ijk} A_i^{\nu} A_j^{\lambda} A_k^{\rho} \right) + \right. \nonumber \\ 
	&& \qquad \qquad \qquad \qquad \left. \sum_{\alpha =3,8} \; \sum_{j,k\, \in \, \mathcal{I}_{\text{off}}} g_s f_{\alpha jk} A_{\alpha}^{\nu} A_j^{\lambda} A_k^{\rho} \; + \, \left( A \rightarrow A_{(d)} \right)\, \right\rbrace \, ,
	\end{eqnarray}
	where $\mathcal{I}_{\text{off}} = [1;7]/ \lbrace 3\rbrace$\footnote{By this notation we mean all integers from 1 to 7 excluding 3.}. Note that each of the interactions presented in Eq.~(\ref{off}) contains two or three off-diagonal (dual) gluon fields. Restricting our study in what follows to the field of low energy QCD, we neglect contribution from these terms. The reason are strong indications~\cite{Suzuki:1989gp,Stack:1994wm,Amemiya:1998jz,Kondo:2014sta} of Abelian dominance in QCD below the energies of $1~\text{GeV}$, which means that the processes involving off-diagonal gluons are suppressed in the IR. Moreover, as we show in Appendix~\ref{appA}, the term~\eqref{off} is exactly zero in the classical approximation. Let us note, however, that in the future it would be very interesting to study if the quantum effects can generate non-zero $\mathcal{L}_{\text{off}}$, because, although such effects are expected to be small in IR, they would be a very distinctive feature of the model we discuss.
	
\begin{figure}[t!]
		\includegraphics[height=12cm]{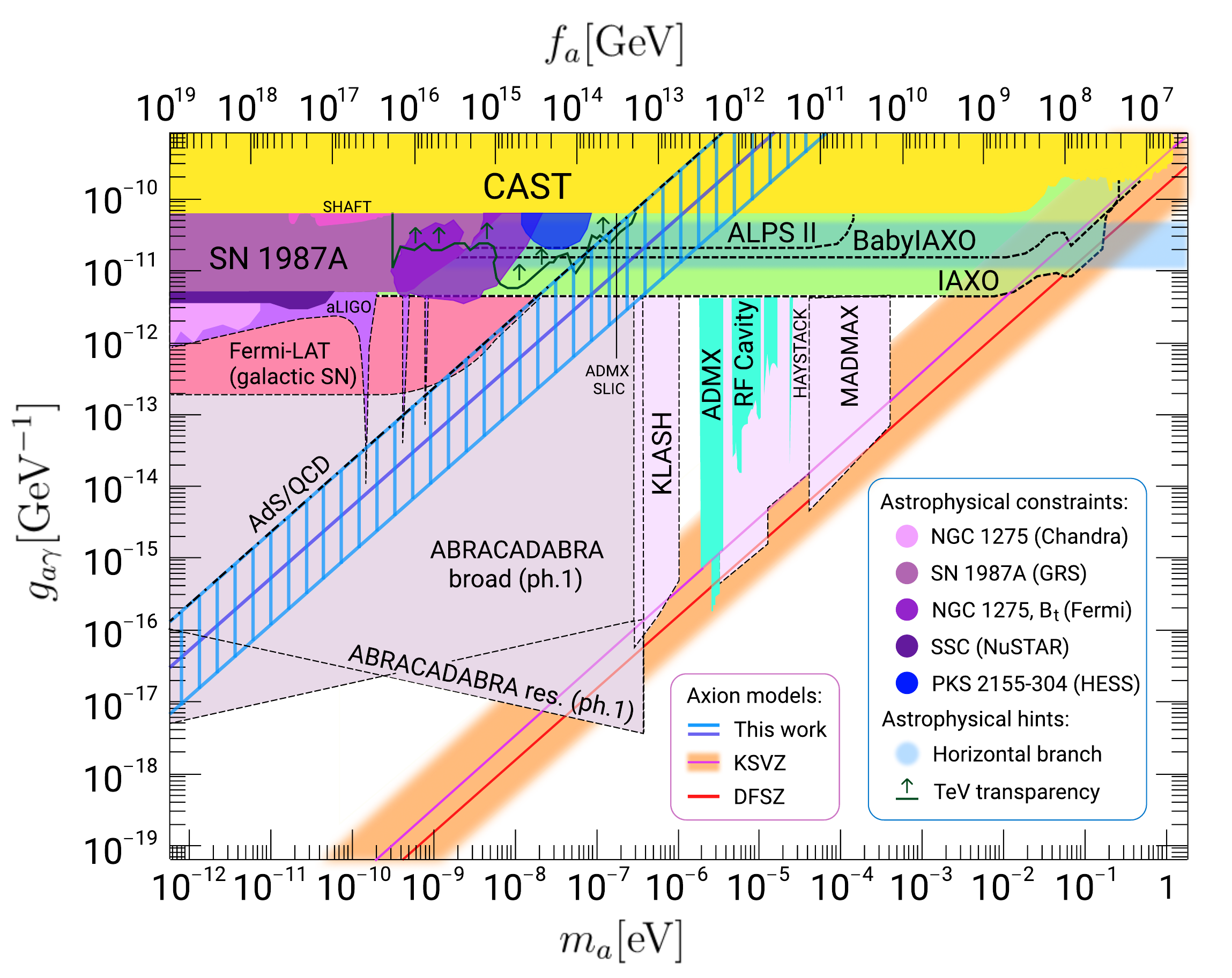}
		\caption{\label{fig1} Axion-photon coupling as a function of axion mass and decay constant for various axion models together with the existing and projected (dashed lines) constraints on the corresponding parameter space from experiments as well as from astrophysical data. Astrophysical hints are also shown. The dash-dotted line corresponds to the model of this work with the non-Abelian monopole where the IR strong coupling $\alpha_s$ value calculated in the AdS/QCD framework is adopted. The line in the center of the vertically hatched band corresponds to the model with the minimal Abelian monopole. For further discussion, see main text.}
	\end{figure}

	The effective Lagrangian Eq.~(\ref{efflag}) without the term $\mathcal{L}_{\text{off}}$ has the form of the conventional axion effective Lagrangian. As we will show, however, the corresponding axion particle has couplings with SM particles which differ a lot from the ones calculated in  DFSZ and KSVZ models. In particular, the coupling to photons $g_{a\gamma}$ is enhanced by many orders of magnitude compared to the conventional models. Namely, after the standard chiral rotation of quarks
	\begin{equation}
	q \rightarrow \exp \left( i\gamma_5\, \frac{a M_q^{-1}}{2f_a \textrm{tr}M_q^{-1}} \right) \cdot q , \quad M_q = \textrm{diag} \left( m_u, m_d \right) ,
	\end{equation} 
	which eliminates the $G \tilde{G}$ term, is performed, one finds that the coupling to photons is
	\begin{equation}\label{gagamma}
	g_{a\gamma} = g^0_{a\gamma} - \frac{\alpha}{3\pi f_a} \left( \frac{4m_d + m_u}{m_d + m_u} \right) \simeq g^0_{a\gamma}, 
	\end{equation}
	so that it is practically not affected by the quark masses. In the conventional notation used to parameterize the strength of the axion-photon coupling,
	\begin{equation}
	g_{a\gamma} = \frac{\alpha}{2\pi f_a} \cdot \frac{E}{N} \, , 
	\end{equation} 
	our model predicts
	\begin{spreadlines}{1ex}
		\begin{equation}
		\frac{E}{N} = \begin{dcases}
		6\alpha_s^2/  \alpha^2\, , \\
		27/ (2\alpha^2)\, ,
		\end{dcases}
		\end{equation}
	\end{spreadlines}
	so that the coupling gets increased by 5-6 orders of magnitude.
	Bearing in mind that the standard expression for the axion mass,
	\begin{equation}\label{mass}
	m_a = \frac{m_{\pi} f_{\pi} \sqrt{m_u m_d}}{\left( m_u + m_d \right) f_a}\, ,
	\end{equation}
	is derived from the conventional axion-gluon coupling and thus holds in our case automatically as long as $\mathcal{L}_{\text{off}}$ is small, we plot the axion-photon coupling as a function of axion mass and decay constant in Fig.~\ref{fig1} together with the hints and existing as well as projected constraints from various experiments and astrophysical observations.\footnote{Hints and most constraints are discussed in detail in Ref.~\cite{DiLuzio:2020wdo}. We present updated astrophysical constraints from Ref.~\cite{2020arXiv200808100A} together with the constraints derived from Chandra data on NGC~1275~\cite{Reynolds_2020} (see however Ref.~\cite{Libanov:2019fzq}), as well as constraints derived from the data on SN1987A from the GRS instrument of the SMM satellite~\cite{Payez:2014xsa}, constraints from NuSTAR data on super star clusters~\cite{Dessert:2020lil} and projected constraints from advanced LIGO~\cite{Nagano:2019rbw}. Constraints from ADMX SLIC~\cite{Crisosto:2019fcj} search for dark matter axions include three very narrow close exclusion regions which are impossible to resolve in our plot. SHAFT constraint is discussed in Ref.~\cite{Gramolin:2020ict}.} For reference, we show axion-photon couplings in KSVZ models with heavy fermions in one representation of the SM gauge group~\cite{DiLuzio:2017pfr} and in DFSZ model. 
	
	In Fig.~\ref{fig1}, possible values for the axion-photon coupling in the model with the non-Abelian monopole are organized in a vertically hatched band, while the model with the minimal Abelian monopole yields a single line inside this band. The band denotes the uncertainty we estimate for the model with the non-Abelian monopole, which is associated to the dependence of the first line of Eq.~(\ref{g0}) on the strong coupling $\alpha_s$ in the IR. The state of the art in studies of the behavior of the latter was discussed in detail in a recent review~\cite{Deur:2016tte}, where it was shown that there exists a definition of $\alpha_s$ in the IR, which is analytic, independent of the choice of renormalization scheme or gauge, universal, based on first principles and IR-finite (see Table 5.4 in Ref.~\cite{Deur:2016tte}). This choice of definition for IR $\alpha_s$ corresponds to the so-called effective charges $\alpha_{g_1}$, $\alpha_{F_3}$ and $\alpha_{\tau}$, which are directly related to the observables of low energy QCD. The measurements show that the IR strong coupling $\alpha_s$ defined in such a way freezes at low energies. The freezing behavior of IR $\alpha_s$ is also supported by the success of the AdS/QCD technique in the description of hadron properties~\cite{Erlich:2005qh}. Moreover, the value of the IR strong coupling calculated in AdS/QCD,  $\alpha_{\text{AdS}} \left( 0 \right) = \pi$, is consistent with the values $\alpha_{g_1} ( 0 ) $ and $\alpha_{F_3} ( 0 ) $ \footnote{Although the effective charge $\alpha_{\tau} ( 0 ) $ is different, it is known that it contains an unsubtracted pion pole.}. All this convinces us to assume that the AdS/QCD value of IR strong coupling is a relevant one, that is why we highlight the corresponding values of $g_{a\gamma}$ in Fig.~\ref{fig1} with a dash-dotted line. However, bearing in mind that low energy QCD is still largely terra incognita, we allow for uncertainty in $\alpha_s$ which results in a band in Fig~\ref{fig1} where the lower edge $\alpha_s (0) = 0.7$ is chosen. Such choice is suggested by the observation in Ref.~\cite{Deur:2016tte} that most of the values of $\alpha_s (0)$ in the literature are clustered around $\alpha_s (0) \sim 3$ (close to the AdS/QCD value) and $\alpha_s (0) \sim 0.7$, not taking into account the decoupling solution  $\alpha_s (0) = 0$ disfavored for a number of reasons~\cite{Fischer:2008uz, Kugo:1979gm}. Let us note as well that too large values of $\alpha_s$ are disfavored by calculations in Ref.~\cite{Liao:2008jg}, where it was shown that the magnetic coupling (i.e. the coupling inverse to $\alpha_s$) never gets too small in pure $SU(2)$ gluodynamics, these results being extended to the pure $SU(3)$ case in Ref.~\cite{Bonati:2013bga}.
	
	Finally, let us mention that there is yet another source of uncertainty in our predictions, both for the models with Abelian and non-Abelian monopoles, which is associated with the $U(1)$ magnetic charges of the monopoles. Whereas we consider them to be minimal in each of the model, they are in principle not constrained by the stability arguments. This means that $g_{a\gamma}$ can be further increased in Fig.~\ref{fig1} for the models of this work.
	
\begin{figure}[t!]
		\includegraphics[height=12cm]{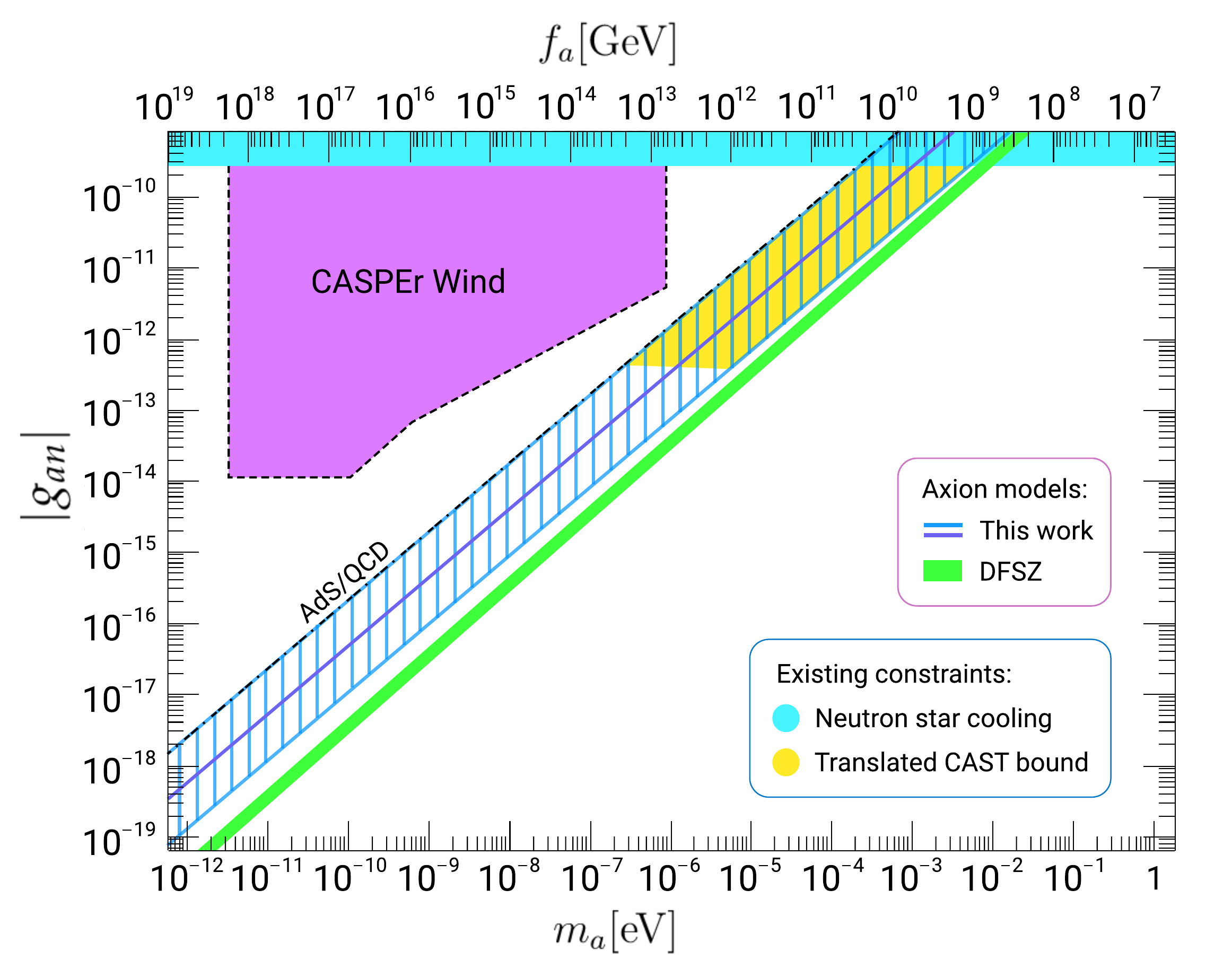}
		\caption{\label{fig2} Axion-neutron coupling as a function of axion mass and decay constant for various axion models together with the existing and projected (dashed lines) constraints on the corresponding parameter space from experiments as well as from astrophysical data. The dash-dotted line corresponds to the model of this work with the non-Abelian monopole where the IR strong coupling $\alpha_s$ value calculated in the AdS/QCD framework is adopted. The line in the center of the vertically hatched band corresponds to the model with the minimal Abelian monopole. For further discussion, see main text.}
	\end{figure}

	Next, let us consider axion couplings with matter, $g_{ai} \equiv C_{ai} m_i/f_a \,$, 
	where $m_i$ is the mass of fermion $i$, 
	which correspond to the following terms in the effective Lagrangian:
	\begin{equation}
	\mathcal{L}_{\text{eff}} \supset C_{ai}\frac{\partial_{\mu} a}{2f_a} \, \bar{\psi}_i \gamma^{\mu} \gamma_5 \psi_i \, .
	\end{equation}
	As electrons do not carry PQ charge in the model we consider, the axion-electron coupling $g_{ae}$ is generated radiatively~\cite{Chang:1993gm,Srednicki:1985xd}:
	\begin{equation}
	g_{ae} = g^0_{a\gamma} \cdot \frac{3\alpha}{2\pi} m_e \ln \frac{f_a}{m_e} \, ,
	\end{equation} 
	where we took into account that the term associated to the axion-pion mixing is negligibly small compared to the leading contribution. We find that the experiments and astrophysical observations probing axion-electron interactions do not yield new constraints on the model. Indeed, the CAST bound~\cite{Anastassopoulos:2017ftl} on the axion-photon coupling, $g_{a\gamma} < 0.66\cdot 10^{-10}~\text{GeV}^{-1}$, constrains the phenomenologically viable region for axion-electron coupling: $g_{ae} < 1.2\cdot 10^{-16} \ln f_a/m_e$. This constraint is stronger than any existing or projected bound from interaction with electrons. As to the interactions of the axion with nucleons, it turns out that contributions from radiatively generated axion-quark couplings are non-negligible and actually enhance axion-nucleon couplings with respect to the conventional DFSZ case in much of the parameter space. One can find that the coefficients $C_{ap}$ and $C_{an}$ are
	\begin{eqnarray}
	\label{cap}&C_{ap} = -0.47 - 0.39\, \delta c_d + 0.88\, \delta c_u \, , \\
	\label{can}&C_{an} = -0.02 + 0.88\, \delta c_d - 0.39\, \delta c_u \, ,
	\end{eqnarray}
	where the numerical coefficients were calculated in~\cite{diCortona:2015ldu} and the radiatively generated quark couplings read as follows:
	\begin{align}
	\label{dcau}&\delta c_u = g^0_{a\gamma} f_a \cdot \frac{8\alpha}{27\pi} \ln \frac{f_a}{m_N} \, , \\[4pt]
	\label{dcad}&\delta c_d = g^0_{a\gamma} f_a \cdot \frac{\alpha}{54\pi} \ln \frac{f_a}{m_N} \, ,
	\end{align}
	where $m_N$ is the nucleon mass. Constraints on axion-neutron interactions are more stringent than constraints on interactions with protons. We plot $g_{an}$ as a function of axion mass and decay constant in Fig.~\ref{fig2} together with the constraint from neutron star cooling~\cite{Beznogov:2018fda} and the projected reach of the CASPEr Wind experiment~\cite{JacksonKimball:2017elr}. For reference, we show the neutron-axion coupling in DFSZ models, the range of which is constrained by the requirement of perturbative unitarity of the Yukawa couplings of SM fermions~\cite{Chen:2013kt}. Note that the slope of the DFSZ band in Fig.~\ref{fig2} is different from the slope of the band corresponding to the axion model of this paper. The difference arises because, in the DFSZ case, one obtains a linear dependence of the coupling on the axion mass, $g_{an} \propto m_a$, characteristic of the tree-level couplings to quarks, while in the case of our model the linear dependence is superseded by a nonlinear one, $g_{an} \propto m_a \ln \left( \mathrm{const} / m_a \right)$, due to the radiative origin of the coupling.
	In Fig.~\ref{fig2}, we show also the CAST bound~\cite{Anastassopoulos:2017ftl} which is translated to a constraint on the axion-neutron coupling with the use of Eqs.~(\ref{can}-\ref{dcad}). Uncertainty in the prediction of the axion-neutron coupling in the axion models of this work comes from the uncertainty in the prediction of the axion-photon coupling, the latter being discussed at length above.
	
	Finally, let us discuss if the axions we propose can comprise dark matter.
	In order to avoid the cosmological magnetic monopole problem~\cite{Zeldovich:1978wj, Preskill:1979zi}, i.e. overproduction of monopoles during the hot Big Bang epoch, we will set their masses and therefore the axion decay constant $f_a$ to be larger than the reheating temperature. This means that we have to deal only with the pre-inflationary scenario of axion dark matter production, which hinges upon the misalignment mechanism~\cite{Abbott:1982af,Dine:1982ah,Preskill:1982cy}. Our model with a heavy Abelian magnetic monopole charged electrically under $SU(3)$ will then give exactly the same dark matter abundance as in the case of the KSVZ model. This follows from the fact that the axion-gluon couplings are identical in the latter two models. Meanwhile, calculation of the dark matter abundance is generally not so simple in the case of our model with a non-Abelian magnetic monopole. Note that while Abelian dominance suggests that the low temperature axion mass $m_a (f_a)$ in this case is given approximately by the familiar expression for the standard QCD axion, at higher temperatures, $T \gtrsim 1~\text{GeV}$, the axion mass can differ significantly from the standard case. The cosmic axion abundance resulting from the misalignment production mechanism $\rho_{a}^{\text{mis}}$  is inversely proportional to the square root of the axion mass at the moment where oscillations of the axion field start:
	\begin{equation}
	\rho_{a}^{\text{mis}} \propto \frac{f_a}{\sqrt{m_a ( T_{\text{roll}} ) }} \cdot \mathcal{F}\, (T_{\text{roll}}) \, ,
	\end{equation}
	where $T_{\text{roll}}$ is the temperature at which $m_a ( T_{\text{roll}} ) = 3H ( T_{\text{roll}} ) $, $H$ being the Hubble expansion rate, and $\mathcal{F}$ a fixed function of temperature. Due to Abelian dominance, we expect that $\rho_{a}^{\text{mis}}$ does not change too much with respect to the conventional QCD axion models if $ T_{\text{roll}} < 1~\text{GeV}$. The latter condition can be recast into the form $m_a (\text{GeV}) < 3H(\text{GeV})$, which yields $f_a > 10^{12}~\text{GeV}$ assuming the axion mass at $1~\text{GeV}$ is not much off the values given in Ref.~\cite{Borsanyi:2016ksw}. Combining it with the CAST bound, we see that in much of the allowed parameter space axions produced via the misalignment mechanism have approximately the same abundance as axions with the same mass in KSVZ and DFSZ-like models. The case $f_a \lesssim 10^{12}~\text{GeV}$ is more difficult: in order to infer the abundance of cosmic axions in the model with a non-Abelian monopole in this case, one has to calculate the axion mass as a function of temperature in the energy range where there is no Abelian dominance. We leave a more thorough investigation of axion cosmology in our model for future work.
	
	\section{Discussion}
	
	In this work, we introduced a new hadronic axion model which involves a very heavy vector-like fermion magnetically charged under either the full non-Abelian symmetry of the low energy SM or only its electromagnetic subgroup. We showed that both cases can realize the PQ mechanism and thus provide a solution to the strong CP problem. We found that both cases lead to a very interesting phenomenology. Although we assumed Abelian dominance in our discussion of phenomenology of the model with the non-Abelian monopole, it is easy to see that the functions $g_{a\gamma} (f_a)$, $g_{ae} (f_a)$ and $g_{an} (f_a)$ are independent of this assumption: the latter two couplings are generated at 1-loop through the coupling to photons $g_{a\gamma}$ while the axion-photon coupling is completely dominated by the $aF\tilde{F}$ term, see Eq.~(\ref{gagamma}). The quantities which are sensitive to the axion interactions with off-diagonal gluons are the axion mass $m_a (f_a)$, the coupling to the nuclear EDM $g_d (f_a)$ and, in some part of the parameter space, the couplings with protons and pions. If there exist non-vanishing quantum corrections to the term $\mathcal{L}_{\text{off}}$ (Eq.~(\ref{off})), the model with the non-Abelian monopole constitutes a counterexample to the assertion of universality of axion-gluon coupling and EDM coupling in QCD axion models. If such corrections are not too small, difference in EDM coupling with respect to the conventional QCD axion models can offer an exciting opportunity of distinguishing our axion model from the other QCD axion models in experiments such as CASPEr Electric~\cite{JacksonKimball:2017elr}.

	It is especially intriguing that the model of the QCD axion we discuss is consistent with the astrophysical hints suggested both by anomalous TeV-transparency of the Universe~\cite{DeAngelis:2008sk, Horns:2012fx} and by excessive cooling of horizontal branch stars in globular clusters~\cite{Ayala:2014pea}, see Fig.~\ref{fig1}. Moreover, Fig.~\ref{fig1} shows that the parameter space of our model is to be probed in the future by many experiments and astrophysical observatories, namely ALPS~II~\cite{Bahre:2013ywa}, \mbox{BabyIAXO}~\cite{Abeln:2020ywv}, IAXO~\cite{Armengaud:2014gea} and Fermi-LAT~\cite{Meyer:2016wrm}. Meanwhile, advanced LIGO~\cite{Nagano:2019rbw}, KLASH~\cite{2017arXiv170706010A}, ADMX SLIC~\cite{Crisosto:2019fcj} and ABRACADABRA~\cite{Kahn:2016aff} experiments have all the chances to discover the cosmic abundance of such axions. As to the experiments which probe the interactions of axion with neutrons, one can see in Fig.~\ref{fig2} that although the projected reach of the CASPEr Wind experiment is not enough to probe the QCD axion model we propose, the gap between theory and experiment is way smaller than in the case of DFSZ axions. 
	
	The model we discussed is peculiar in yet another way. Suppose that the axion is found through its coupling with photons and that investigation of its EDM coupling shows preference for the model involving non-Abelian monopole. Then one can infer the IR strong coupling $\alpha_s (0)$ by Eq.~(\ref{g0}). This would be an independent experimental hint for the coupling constant $\alpha_s (0)$ which could refine other determinations. 
	
	Needless to say, it would be valuable to construct a UV completion to the model discussed. Presence of heavy magnetic monopoles in the spectrum of the UV theory can influence Z-boson physics, possibly providing an additional opportunity for probing the model of this work experimentally. An interesting question regarding the UV completion is whether the magnetic charges we discuss can emerge from some Grand Unified theory via the 't Hooft-Polyakov construction~\cite{tHooft:1974kcl, Polyakov:1974ek}. Note that the model of this work requires magnetic charges to be carried by fermionic particles. The latter can arise as systems of magnetically and electrically charged bosons~\cite{Goldhaber:1976dp}, e.g. as pairs of identical dyons. Fermionic monopoles naturally arise in supersymmetric theories.
	
	\medskip
	
	\appendix
	\section{Axion-gluon coupling in the classical approximation}\label{appA}
	In this Appendix, we show that the axion-gluon coupling in the model with a heavy non-Abelian monopole preserves its universality in the classical approximation, i.e. it is given by the expression:
	\begin{equation}\label{agg}
	-\, \frac{a g_s^2}{32\pi^2 f_a}\; G_{\mu \nu}^a \tilde{G}^{a\, \mu \nu},
	\end{equation}
	so that $\mathcal{L}_{\text{off}} = 0\, $ in Eq.~(\ref{efflag}), at least classicaly. We use the formalism of loop space variables pioneered by Polyakov~\cite{Polyakov:1980ca} and developed with the focus on the electric-magnetic dual symmetry of the YM theory by Hong-Mo, Faridani and Tsun~\cite{Chan:1995xr}. Central object of the formalism is the parallel phase transport along the loop $\xi (s), \, s \in \left[ 0, 2\pi \right]$ from one point $s_1$ to another $s_2$:
	\begin{equation}
	\Phi_{\xi} (s_2,s_1) = P_s \exp{\left( ig_s \int_{s_1}^{s_2} ds \, A_{\mu } \left( \xi (s) \right) \dot{\xi}^{\mu} (s) \right)},
	\end{equation}
	where $P_s$ is the Dyson ordering. Loop derivative of the holonomy defines the Polyakov variables:
	\begin{equation}
	F_{\mu}[\xi|s] = \frac{i}{g_s} \Phi_{\xi}^{-1} \left( 2\pi, 0 \right) \cdot \frac{\delta \Phi_{\xi} \left( 2\pi, 0 \right) }{\delta \xi^{\mu} (s)} \, ,
	\end{equation}
	which are known to constitute a valid set for a full description of the YM field~\cite{Chan:1985ix, Chan:1985xc}. It was shown in Ref.~\cite{Chan:1995xr} that another complete set of variables is better suited for dealing with the electric-magnetic dual symmetry of the classical YM theory, namely:
	\begin{equation}
	E_{\mu} \left[ \xi | s \right] = \Phi_{\xi} \left( s, 0 \right) F_{\mu} \left[ \xi | s \right] \Phi_{\xi}^{-1} \left( s, 0 \right) ,
	\end{equation}
	which can be connected to the local quantities by the expression:
	\begin{equation}\label{locloop}
	\omega^{-1} \! \left( x \right) \, \tilde{G}_{\mu \nu} \, (x) \; \omega \left( x \right) = \frac{2}{N}\, \epsilon_{\mu \nu \rho \sigma} \int \delta \xi ds \,  E^{\rho} \left[ \xi | s \right] \dot{\xi}^{\sigma} (s)\, \dot{\xi}^{-2} (s) \, \delta (x-\xi (s)) \, ,
	\end{equation}
	where $\omega \left( x \right)$ is an arbitrary local $SU(3)$ matrix and $N$ is a normalization factor. The dual (magnetic) variables $E^{(d)}_{\mu}$ were shown to be related to the electric ones $E_{\mu}$ in the pure YM theory in the following way:
	\begin{equation}\label{dualtransf}
	\omega^{-1} \! \left( \eta (t) \right) E^{(d)}_{\mu} \left[ \eta | t \right] \, \omega \left( \eta (t) \right) = \frac{2}{N}\, \epsilon_{\mu \nu \rho \sigma} \dot{\eta}^{\nu} (t) \int \delta \xi ds \, E^{\rho} \left[ \xi | s \right] \dot{\xi}^{\sigma} (s)\, \dot{\xi}^{-2} (s) \, \delta (\xi (s) - \eta (t)) \, ,
	\end{equation}
	while the inverse transformation is:
	\begin{equation}\label{dualtransfinv}
	\omega \! \left( \eta (t) \right) E_{\mu} \left[ \eta | t \right] \, \omega^{-1} \left( \eta (t) \right) = - \frac{2}{N}\, \epsilon_{\mu \nu \rho \sigma} \dot{\eta}^{\nu} (t) \int \delta \xi ds \, E^{(d)\, \rho} \left[ \xi | s \right] \dot{\xi}^{\sigma} (s)\, \dot{\xi}^{-2} (s) \, \delta (\xi (s) - \eta (t)) \, .
	\end{equation}
	Since in the derivation of the axion effective Lagrangian external fields can be considered constant and homogeneous, as discussed in Sec.~\ref{sec4}, we can apply Eqs.~(\ref{dualtransf}) and~(\ref{dualtransfinv}) in order to find the relation between the expression~\ref{agg} and its dual analogue, constructed from the GNO group connection, in the classical theory. The calculation proceeds as follows:
	\begin{eqnarray}\label{long}
	&&\int d^4 x \, a\! \left( x \right) G_{\mu \nu}^a \! \left( x \right) \tilde{G}^{a\, \mu \nu} \! \left( x \right) = 2	\int d^4 x \, a\! \left( x \right) \textrm{tr} \left\lbrace \omega^{-1} \! \left( x \right)  G_{\mu \nu} \! \left( x \right) \omega \left( x \right) \omega^{-1} \! \left( x \right) \, \tilde{G}_{\mu \nu} \, (x) \; \omega \left( x \right) \right\rbrace =\nonumber \\ 
	&&\qquad \frac{8}{N}\int d^4 x\, \delta \xi ds\,  a\! \left( x \right) \textrm{tr} \left\lbrace \omega^{-1} \! \left( x \right) \, \tilde{G}_{\mu \nu} \, (x) \; \omega \left( x \right) E^{\mu} \left[ \xi | s \right] \right\rbrace   \dot{\xi}^{\nu} (s)\, \dot{\xi}^{-2} (s) \, \delta (x - \xi (s)) = \nonumber \\
	&& \frac{16}{N^2}\, \epsilon_{\mu \nu \rho \sigma}\! \int \delta \eta dt\, \delta \xi ds\, a\! \left( \eta (t) \right) \textrm{tr} \left\lbrace E^{\rho} \left[ \eta | t \right]  E^{\mu} \left[ \xi | s \right] \right\rbrace \dot{\eta}^{\sigma} (t)\, \dot{\eta}^{-2} (t) \, \dot{\xi}^{\nu} (s)\, \dot{\xi}^{-2} (s) \, \delta (\eta (t) - \xi (s)) = \qquad \nonumber \\
	&&\qquad \frac{8}{N} \int \delta \eta dt\, a\! \left( \eta (t) \right) \textrm{tr} \left\lbrace E^{\mu} \left[ \eta | t \right] \, \omega^{-1} \! \left( \eta (t) \right) E^{(d)}_{\mu} \left[ \eta | t \right] \, \omega \left( \eta (t) \right)  \right\rbrace \dot{\eta}^{-2} (t) = \nonumber \\
	&& \frac{8}{N} \int \delta \eta dt\, a\! \left( \eta (t) \right) \textrm{tr} \left\lbrace \, \omega \left( \eta (t) \right)  E^{\mu} \left[ \eta | t \right] \, \omega^{-1} \! \left( \eta (t) \right) E_{\mu}^{(d)} \left[ \eta | t \right] \,  \right\rbrace \dot{\eta}^{-2} (t) = \nonumber \\
	&& - \frac{16}{N^2}\, \epsilon^{\mu \nu \rho \sigma}\! \int \delta \eta dt\, \delta \xi ds\, a\! \left( \eta (t) \right) \textrm{tr} \left\lbrace E^{(d)}_{\rho} \left[ \eta | t \right]  E^{(d)}_{\mu} \left[ \xi | s \right] \right\rbrace \dot{\eta}^{\sigma} (t)\, \dot{\eta}^{-2} (t) \, \dot{\xi}^{\nu} (s)\, \dot{\xi}^{-2} (s) \, \delta (\eta (t) - \xi (s)) = \qquad \nonumber \\
	&&\quad - \int d^4 x \, a\! \left( x \right) G^a_{(d)}{{}_{\mu \nu}} \! \left( x \right) \tilde{G}_{(d)}^{a\, \mu \nu} \! \left( x \right)
	\end{eqnarray}
	where we took advantage of Eqs.~(\ref{locloop}),~(\ref{dualtransf}) and~\ref{dualtransfinv}, as well as of the cyclic property of the trace. The last identity follows automatically as far as one notices that the third and the sixth lines of the Eq.~(\ref{long}) are identical but for the overall sign and electric-magnetic variables interchange. Now, one can clearly see that classically we recover the universal axion-gluon coupling even in the model with the non-Abelian magnetic monopole:
	\begin{equation}
	\mathcal{S}_{\text{eff, classical}} \; \supset \; \int d^4 x\; \frac{a g_s^2}{32\pi^2 f_a}\; G_{(d)}^a{{}_{\mu \nu}} \tilde{G}_{(d)}^{a\, \mu \nu} =  -\int d^4 x\; \frac{a g_s^2}{32\pi^2 f_a}\; G_{\mu \nu}^a \tilde{G}^{a\, \mu \nu} \, .
	\end{equation}
	
	\medskip
	
	\section*{Acknowlegments}
	We thank Claudio Bonati and Thomas Biek{\"o}tter for discussions. 
	A.R. acknowledges support and A.S. is funded by the Deutsche Forschungsgemeinschaft (DFG, German Research Foundation) under Germany's Excellence Strategy -- EXC 2121 \textit{Quantum Universe} -- 390833306.
	
	\bibliography{maxion}

\begin{thebibliography}{10}

\bibitem{Abel:2020gbr}
C.~Abel et~al.
\newblock {Measurement of the permanent electric dipole moment of the neutron}.
\newblock {\em Phys. Rev. Lett.}, 124(8):081803, 2020, 2001.11966.

\bibitem{Peccei:1977hh}
R.D. Peccei and Helen~R. Quinn.
\newblock {CP Conservation in the Presence of Instantons}.
\newblock {\em Phys. Rev. Lett.}, 38:1440--1443, 1977.

\bibitem{Peccei:1977ur}
R.D. Peccei and Helen~R. Quinn.
\newblock {Constraints Imposed by CP Conservation in the Presence of
  Instantons}.
\newblock {\em Phys. Rev. D}, 16:1791--1797, 1977.

\bibitem{Weinberg:1977ma}
Steven Weinberg.
\newblock {A New Light Boson?}
\newblock {\em Phys. Rev. Lett.}, 40:223--226, 1978.

\bibitem{Wilczek:1977pj}
Frank Wilczek.
\newblock {Problem of Strong $P$ and $T$ Invariance in the Presence of
  Instantons}.
\newblock {\em Phys. Rev. Lett.}, 40:279--282, 1978.

\bibitem{Preskill:1982cy}
John Preskill, Mark~B. Wise, and Frank Wilczek.
\newblock {Cosmology of the Invisible Axion}.
\newblock {\em Phys. Lett. B}, 120:127--132, 1983.

\bibitem{Abbott:1982af}
L.F. Abbott and P.~Sikivie.
\newblock {A Cosmological Bound on the Invisible Axion}.
\newblock {\em Phys. Lett. B}, 120:133--136, 1983.

\bibitem{Dine:1982ah}
Michael Dine and Willy Fischler.
\newblock {The Not So Harmless Axion}.
\newblock {\em Phys. Lett. B}, 120:137--141, 1983.

\bibitem{Kim:1979if}
Jihn~E. Kim.
\newblock {Weak Interaction Singlet and Strong CP Invariance}.
\newblock {\em Phys. Rev. Lett.}, 43:103, 1979.

\bibitem{Shifman:1979if}
Mikhail~A. Shifman, A.I. Vainshtein, and Valentin~I. Zakharov.
\newblock {Can Confinement Ensure Natural CP Invariance of Strong
  Interactions?}
\newblock {\em Nucl. Phys. B}, 166:493--506, 1980.

\bibitem{Dine:1981rt}
Michael Dine, Willy Fischler, and Mark Srednicki.
\newblock {A Simple Solution to the Strong CP Problem with a Harmless Axion}.
\newblock {\em Phys. Lett. B}, 104:199--202, 1981.

\bibitem{Zhitnitsky:1980tq}
A.R. Zhitnitsky.
\newblock {On Possible Suppression of the Axion Hadron Interactions. (In
  Russian)}.
\newblock {\em Sov. J. Nucl. Phys.}, 31:260, 1980.

\bibitem{Farina:2016tgd}
Marco Farina, Duccio Pappadopulo, Fabrizio Rompineve, and Andrea Tesi.
\newblock {The photo-philic QCD axion}.
\newblock {\em JHEP}, 01:095, 2017, 1611.09855.

\bibitem{Hook:2018jle}
Anson Hook.
\newblock {Solving the Hierarchy Problem Discretely}.
\newblock {\em Phys. Rev. Lett.}, 120(26):261802, 2018, 1802.10093.

\bibitem{DiLuzio:2021pxd}
Luca Di~Luzio, Belen Gavela, Pablo Quilez, and Andreas Ringwald.
\newblock {An even lighter QCD axion}.
\newblock {\em arXiv e-prints}, Jan 2021, 2102.00012.

\bibitem{Ayala:2014pea}
Adrian Ayala, Inma Dom\'\i{}nguez, Maurizio Giannotti, Alessandro Mirizzi, and
  Oscar Straniero.
\newblock {Revisiting the bound on axion-photon coupling from Globular
  Clusters}.
\newblock {\em Phys. Rev. Lett.}, 113(19):191302, 2014, 1406.6053.

\bibitem{DeAngelis:2008sk}
A.~De~Angelis, O.~Mansutti, M.~Persic, and M.~Roncadelli.
\newblock {Photon propagation and the VHE gamma-ray spectra of blazars: how
  transparent is really the Universe?}
\newblock {\em Mon. Not. Roy. Astron. Soc.}, 394:L21--L25, 2009, 0807.4246.

\bibitem{Horns:2012fx}
D.~Horns and M.~Meyer.
\newblock {Indications for a pair-production anomaly from the propagation of
  VHE gamma-rays}.
\newblock {\em JCAP}, 02:033, 2012, 1201.4711.

\bibitem{Dirac:1931kp}
Paul Adrien~Maurice Dirac.
\newblock {Quantised singularities in the electromagnetic field,}.
\newblock {\em Proc. Roy. Soc. Lond. A}, 133(821):60--72, 1931.

\bibitem{PhysRevD.3.880}
Daniel Zwanziger.
\newblock Local-lagrangian quantum field theory of electric and magnetic
  charges.
\newblock {\em Phys. Rev. D}, 3:880--891, Feb 1971.

\bibitem{PhysRevLett.40.147}
Richard~A. Brandt, Filippo Neri, and Daniel Zwanziger.
\newblock Lorentz invariance of the quantum field theory of electric and
  magnetic charge.
\newblock {\em Phys. Rev. Lett.}, 40:147--150, Jan 1978.

\bibitem{PhysRevD.19.1153}
Richard~A. Brandt, Filippo Neri, and Daniel Zwanziger.
\newblock Lorentz invariance from classical particle paths in quantum field
  theory of electric and magnetic charge.
\newblock {\em Phys. Rev. D}, 19:1153--1167, Feb 1979.

\bibitem{Rubakov:1981rg}
V.~A. Rubakov.
\newblock {Superheavy Magnetic Monopoles and Proton Decay}.
\newblock {\em JETP Lett.}, 33:644--646, 1981.

\bibitem{Callan:1982au}
Curtis~G. Callan, Jr.
\newblock {Dyon-Fermion Dynamics}.
\newblock {\em Phys. Rev. D}, 26:2058--2068, 1982.

\bibitem{Englert:1976ng}
F.~Englert and Paul Windey.
\newblock {Quantization Condition for 't Hooft Monopoles in Compact Simple Lie
  Groups}.
\newblock {\em Phys. Rev. D}, 14:2728, 1976.

\bibitem{Goddard:1976qe}
P.~Goddard, J.~Nuyts, and David~I. Olive.
\newblock {Gauge Theories and Magnetic Charge}.
\newblock {\em Nucl. Phys. B}, 125:1--28, 1977.

\bibitem{Montonen:1977sn}
C.~Montonen and David~I. Olive.
\newblock {Magnetic Monopoles as Gauge Particles?}
\newblock {\em Phys. Lett. B}, 72:117--120, 1977.

\bibitem{Kapustin:2006pk}
Anton Kapustin and Edward Witten.
\newblock {Electric-Magnetic Duality And The Geometric Langlands Program}.
\newblock {\em Commun. Num. Theor. Phys.}, 1:1--236, 2007, hep-th/0604151.

\bibitem{Chan:1995xr}
Hong-Mo Chan, J.~Faridani, and Sheung-Tsun Tsou.
\newblock {A Generalized duality symmetry for nonAbelian Yang-Mills fields}.
\newblock {\em Phys. Rev. D}, 53:7293--7305, 1996, hep-th/9512173.

\bibitem{Wu:1975es}
Tai~Tsun Wu and Chen~Ning Yang.
\newblock {Concept of Nonintegrable Phase Factors and Global Formulation of
  Gauge Fields}.
\newblock {\em Phys. Rev. D}, 12:3845--3857, 1975.

\bibitem{Brandt:1979kk}
Richard~A. Brandt and Filippo Neri.
\newblock {Stability Analysis for Singular Nonabelian Magnetic Monopoles}.
\newblock {\em Nucl. Phys. B}, 161:253--282, 1979.

\bibitem{Coleman:1982cx}
Sidney~R. Coleman.
\newblock {THE MAGNETIC MONOPOLE FIFTY YEARS LATER}.
\newblock In {\em {Les Houches Summer School of Theoretical Physics:
  Laser-Plasma Interactions}}, 6 1982.

\bibitem{tHooft:1981bkw}
Gerard 't~Hooft.
\newblock {Topology of the Gauge Condition and New Confinement Phases in
  Nonabelian Gauge Theories}.
\newblock {\em Nucl. Phys. B}, 190:455--478, 1981.

\bibitem{Bonati:2010tz}
Claudio Bonati, Adriano Di~Giacomo, Luca Lepori, and Fabrizio Pucci.
\newblock {Monopoles, abelian projection and gauge invariance}.
\newblock {\em Phys. Rev. D}, 81:085022, 2010, 1002.3874.

\bibitem{Kondo:2014sta}
Kei-Ichi Kondo, Seikou Kato, Akihiro Shibata, and Toru Shinohara.
\newblock {Quark confinement: Dual superconductor picture based on a
  non-Abelian Stokes theorem and reformulations of Yang-Mills theory}.
\newblock {\em Phys. Rept.}, 579:1--226, 2015, 1409.1599.

\bibitem{Amemiya:1998jz}
Kazuhisa Amemiya and Hideo Suganuma.
\newblock {Off diagonal gluon mass generation and infrared Abelian dominance in
  the maximally Abelian gauge in lattice QCD}.
\newblock {\em Phys. Rev. D}, 60:114509, 1999, hep-lat/9811035.

\bibitem{Suzuki:2017lco}
Tsuneo Suzuki, Katsuya Ishiguro, and Vitaly Bornyakov.
\newblock {New scheme for color confinement and violation of the non-Abelian
  Bianchi identities}.
\newblock {\em Phys. Rev. D}, 97(3):034501, 2018, 1712.05941.
\newblock [Erratum: Phys.Rev.D 97, 099905 (2018)].

\bibitem{Witten:1979ey}
Edward Witten.
\newblock {Dyons of Charge e theta/2 pi}.
\newblock {\em Phys. Lett. B}, 86:283--287, 1979.

\bibitem{Vafa:1984xg}
Cumrun Vafa and Edward Witten.
\newblock {Parity Conservation in QCD}.
\newblock {\em Phys. Rev. Lett.}, 53:535, 1984.

\bibitem{Schwinger:1951nm}
Julian~S. Schwinger.
\newblock {On gauge invariance and vacuum polarization}.
\newblock {\em Phys. Rev.}, 82:664--679, 1951.

\bibitem{Suzuki:1989gp}
Tsuneo Suzuki and Ichiro Yotsuyanagi.
\newblock {A possible evidence for Abelian dominance in quark confinement}.
\newblock {\em Phys. Rev. D}, 42:4257--4260, 1990.

\bibitem{Stack:1994wm}
John~D. Stack, Steven~D. Neiman, and Roy~J. Wensley.
\newblock {String tension from monopoles in SU(2) lattice gauge theory}.
\newblock {\em Phys. Rev. D}, 50:3399--3405, 1994, hep-lat/9404014.

\bibitem{DiLuzio:2020wdo}
Luca Di~Luzio, Maurizio Giannotti, Enrico Nardi, and Luca Visinelli.
\newblock {The landscape of QCD axion models}.
\newblock {\em Phys. Rept.}, 870:1--117, 2020, 2003.01100.

\bibitem{2020arXiv200808100A}
Gautham {Adamane Pallathadka}, Francesca {Calore}, Pierluca {Carenza}, Maurizio
  {Giannotti}, Dieter {Horns}, Jhilik {Majumdar}, Alessandro {Mirizzi}, Andreas
  {Ringwald}, Anton {Sokolov}, and Franziska {Stief}.
\newblock {Reconciling hints on axion-like-particles from high-energy gamma
  rays with stellar bounds}.
\newblock {\em arXiv e-prints}, Aug 2020, 2008.08100.

\bibitem{Reynolds_2020}
Christopher~S. Reynolds, M.~C.~David Marsh, Helen~R. Russell, Andrew~C. Fabian,
  Robyn Smith, Francesco Tombesi, and Sylvain Veilleux.
\newblock {Astrophysical Limits on Very Light Axion-like Particles from Chandra
  Grating Spectroscopy of NGC 1275}.
\newblock {\em The Astrophysical Journal}, 890(1):59, 2020, 1907.05475.

\bibitem{Libanov:2019fzq}
Maxim Libanov and Sergey Troitsky.
\newblock {On the impact of magnetic-field models in galaxy clusters on
  constraints on axion-like particles from the lack of irregularities in
  high-energy spectra of astrophysical sources}.
\newblock {\em Phys. Lett. B}, 802:135252, 2020, 1908.03084.

\bibitem{Payez:2014xsa}
Alexandre Payez, Carmelo Evoli, Tobias Fischer, Maurizio Giannotti, Alessandro
  Mirizzi, and Andreas Ringwald.
\newblock {Revisiting the SN1987A gamma-ray limit on ultralight axion-like
  particles}.
\newblock {\em JCAP}, 02:006, 2015, 1410.3747.

\bibitem{Dessert:2020lil}
Christopher Dessert, Joshua~W. Foster, and Benjamin~R. Safdi.
\newblock {X-ray Searches for Axions from Super Star Clusters}.
\newblock {\em Phys. Rev. Lett.}, 125(26):261102, 2020, 2008.03305.

\bibitem{Nagano:2019rbw}
Koji Nagano, Tomohiro Fujita, Yuta Michimura, and Ippei Obata.
\newblock {Axion Dark Matter Search with Interferometric Gravitational Wave
  Detectors}.
\newblock {\em Phys. Rev. Lett.}, 123(11):111301, 2019, 1903.02017.

\bibitem{Crisosto:2019fcj}
N.~Crisosto, P.~Sikivie, N.~S. Sullivan, D.~B. Tanner, J.~Yang, and G.~Rybka.
\newblock {ADMX SLIC: Results from a Superconducting $LC$ Circuit Investigating
  Cold Axions}.
\newblock {\em Phys. Rev. Lett.}, 124(24):241101, 2020, 1911.05772.

\bibitem{Gramolin:2020ict}
Alexander~V. Gramolin, Deniz Aybas, Dorian Johnson, Janos Adam, and
  Alexander~O. Sushkov.
\newblock {Search for axion-like dark matter with ferromagnets}.
\newblock {\em Nature Phys.}, 17(1):79--84, 2021, 2003.03348.

\bibitem{DiLuzio:2017pfr}
Luca Di~Luzio, Federico Mescia, and Enrico Nardi.
\newblock {Window for preferred axion models}.
\newblock {\em Phys. Rev. D}, 96(7):075003, 2017, 1705.05370.

\bibitem{Deur:2016tte}
Alexandre Deur, Stanley~J. Brodsky, and Guy~F. de~Teramond.
\newblock {The QCD Running Coupling}.
\newblock {\em Nucl. Phys.}, 90:1, 2016, 1604.08082.

\bibitem{Erlich:2005qh}
Joshua Erlich, Emanuel Katz, Dam~T. Son, and Mikhail~A. Stephanov.
\newblock {QCD and a holographic model of hadrons}.
\newblock {\em Phys. Rev. Lett.}, 95:261602, 2005, hep-ph/0501128.

\bibitem{Fischer:2008uz}
Christian~S. Fischer, Axel Maas, and Jan~M. Pawlowski.
\newblock {On the infrared behavior of Landau gauge Yang-Mills theory}.
\newblock {\em Annals Phys.}, 324:2408--2437, 2009, 0810.1987.

\bibitem{Kugo:1979gm}
Taichiro Kugo and Izumi Ojima.
\newblock {Local Covariant Operator Formalism of Nonabelian Gauge Theories and
  Quark Confinement Problem}.
\newblock {\em Prog. Theor. Phys. Suppl.}, 66:1--130, 1979.

\bibitem{Liao:2008jg}
Jinfeng Liao and Edward Shuryak.
\newblock {Magnetic Component of Quark-Gluon Plasma is also a Liquid!}
\newblock {\em Phys. Rev. Lett.}, 101:162302, 2008, 0804.0255.

\bibitem{Bonati:2013bga}
Claudio Bonati and Massimo D'Elia.
\newblock {The Maximal Abelian Gauge in SU(N) gauge theories and thermal
  monopoles for N = 3}.
\newblock {\em Nucl. Phys. B}, 877:233--259, 2013, 1308.0302.

\bibitem{Chang:1993gm}
Sanghyeon Chang and Kiwoon Choi.
\newblock {Hadronic axion window and the big bang nucleosynthesis}.
\newblock {\em Phys. Lett. B}, 316:51--56, 1993, hep-ph/9306216.

\bibitem{Srednicki:1985xd}
Mark Srednicki.
\newblock {Axion Couplings to Matter. 1. CP Conserving Parts}.
\newblock {\em Nucl. Phys. B}, 260:689--700, 1985.

\bibitem{Anastassopoulos:2017ftl}
V.~Anastassopoulos et~al.
\newblock {New CAST Limit on the Axion-Photon Interaction}.
\newblock {\em Nature Phys.}, 13:584--590, 2017, 1705.02290.

\bibitem{diCortona:2015ldu}
Giovanni Grilli~di Cortona, Edward Hardy, Javier Pardo~Vega, and Giovanni
  Villadoro.
\newblock {The QCD axion, precisely}.
\newblock {\em JHEP}, 01:034, 2016, 1511.02867.

\bibitem{Beznogov:2018fda}
Mikhail~V. Beznogov, Ermal Rrapaj, Dany Page, and Sanjay Reddy.
\newblock {Constraints on Axion-like Particles and Nucleon Pairing in Dense
  Matter from the Hot Neutron Star in HESS J1731-347}.
\newblock {\em Phys. Rev. C}, 98(3):035802, 2018, 1806.07991.

\bibitem{JacksonKimball:2017elr}
D.F. Jackson~Kimball et~al.
\newblock {Overview of the Cosmic Axion Spin Precession Experiment (CASPEr)}.
\newblock {\em Springer Proc. Phys.}, 245:105--121, 2020, 1711.08999.

\bibitem{Chen:2013kt}
Chien-Yi Chen and S.~Dawson.
\newblock {Exploring Two Higgs Doublet Models Through Higgs Production}.
\newblock {\em Phys. Rev. D}, 87:055016, 2013, 1301.0309.

\bibitem{Zeldovich:1978wj}
Ya.~B. Zeldovich and M.~Yu. Khlopov.
\newblock {On the Concentration of Relic Magnetic Monopoles in the Universe}.
\newblock {\em Phys. Lett. B}, 79:239--241, 1978.

\bibitem{Preskill:1979zi}
John Preskill.
\newblock {Cosmological Production of Superheavy Magnetic Monopoles}.
\newblock {\em Phys. Rev. Lett.}, 43:1365, 1979.

\bibitem{Borsanyi:2016ksw}
Sz. Borsanyi et~al.
\newblock {Calculation of the axion mass based on high-temperature lattice
  quantum chromodynamics}.
\newblock {\em Nature}, 539(7627):69--71, 2016, 1606.07494.

\bibitem{Bahre:2013ywa}
Robin B\"ahre et~al.
\newblock {Any light particle search II \textemdash{}Technical Design Report}.
\newblock {\em JINST}, 8:T09001, 2013, 1302.5647.

\bibitem{Abeln:2020ywv}
A.~Abeln et~al.
\newblock {Conceptual Design of BabyIAXO, the intermediate stage towards the
  International Axion Observatory}.
\newblock Oct 2020, 2010.12076.

\bibitem{Armengaud:2014gea}
E.~Armengaud et~al.
\newblock {Conceptual Design of the International Axion Observatory (IAXO)}.
\newblock {\em JINST}, 9:T05002, 2014, 1401.3233.

\bibitem{Meyer:2016wrm}
M.~Meyer, M.~Giannotti, A.~Mirizzi, J.~Conrad, and M.A. S\'anchez-Conde.
\newblock {Fermi Large Area Telescope as a Galactic Supernovae Axionscope}.
\newblock {\em Phys. Rev. Lett.}, 118(1):011103, 2017, 1609.02350.

\bibitem{2017arXiv170706010A}
David {Alesini}, Danilo {Babusci}, Daniele {Di Gioacchino}, Claudio {Gatti},
  Gianluca {Lamanna}, and Carlo {Ligi}.
\newblock {The KLASH Proposal}.
\newblock {\em arXiv e-prints}, July 2017, 1707.06010.

\bibitem{Kahn:2016aff}
Yonatan Kahn, Benjamin~R. Safdi, and Jesse Thaler.
\newblock {Broadband and Resonant Approaches to Axion Dark Matter Detection}.
\newblock {\em Phys. Rev. Lett.}, 117(14):141801, 2016, 1602.01086.

\bibitem{tHooft:1974kcl}
Gerard 't~Hooft.
\newblock {Magnetic Monopoles in Unified Gauge Theories}.
\newblock {\em Nucl. Phys. B}, 79:276--284, 1974.

\bibitem{Polyakov:1974ek}
Alexander~M. Polyakov.
\newblock {Particle Spectrum in the Quantum Field Theory}.
\newblock {\em JETP Lett.}, 20:194--195, 1974.

\bibitem{Goldhaber:1976dp}
Alfred~S. Goldhaber.
\newblock {Spin and Statistics Connection for Charge-Monopole Composites}.
\newblock {\em Phys. Rev. Lett.}, 36:1122--1125, 1976.

\bibitem{Polyakov:1980ca}
Alexander~M. Polyakov.
\newblock {Gauge Fields as Rings of Glue}.
\newblock {\em Nucl. Phys. B}, 164:171--188, 1980.

\bibitem{Chan:1985ix}
Hong-Mo Chan and Sheung~Tsun Tsou.
\newblock {Gauge Theories in Loop Space}.
\newblock {\em Acta Phys. Polon. B}, 17:259, 1986.

\bibitem{Chan:1985xc}
Hong-Mo Chan, Peter Scharbach, and Sheung~Tsun Tsou.
\newblock {On Loop Space Formulation of Gauge Theories}.
\newblock {\em Annals Phys.}, 166:396--421, 1986.

\end{thebibliography}
	
\end{document}